\documentclass[preprint]{aastex}

\newcommand{\ms}{\mbox{m\,s$^{-1}$}}

\newcommand{\eboo}{\mbox{$\eta$~Boo}}

\newcommand{\Dnu}[1]{\Delta \nu_{#1}}
\newcommand{\dnu}[1]{\delta \nu_{#1}}
\newcommand{\half}{{\textstyle\frac{1}{2}}}
\newcommand{\sixth}{{\textstyle\frac{1}{6}}}
\newcommand{\muHz}{\mbox{$\mu$Hz}}
\let\epsilon\varepsilon

\slugcomment{To appear in AJ (September 2003)}

\shorttitle{Oscillations in $\eta$ Boo}
\shortauthors{Kjeldsen et al.}

\begin{document}

\title{Confirmation of solar-like oscillations in eta Bootis}

\author{
H.    Kjeldsen,\altaffilmark{1,2}
T. R. Bedding,\altaffilmark{3}
I. K. Baldry,\altaffilmark{3,4}
H.    Bruntt,\altaffilmark{2}
R. P. Butler,\altaffilmark{5}
D. A. Fischer,\altaffilmark{6}\\
S.    Frandsen,\altaffilmark{2}
E. L. Gates,\altaffilmark{6}
F.    Grundahl,\altaffilmark{2}
K.    Lang,\altaffilmark{1}
G. W. Marcy,\altaffilmark{6}
A.    Misch,\altaffilmark{7} and
S. S. Vogt\altaffilmark{7} 
}

\altaffiltext{1}{Teoretisk Astrofysik Center, Danmarks Grundforskningsfond,
8000 Aarhus C, Denmark; {\tt hans@phys.au.dk}}

\altaffiltext{2}{Institut for Fysik og Astronomi, 8000 Aarhus C, Denmark}

\altaffiltext{3}{School of Physics A28, University of Sydney, NSW 2006,
Australia}

\altaffiltext{4}{Department of Physics and Astronomy, Johns Hopkins
University, Baltimore, MD 21218-2686, USA}

\altaffiltext{5}{Department of Terrestrial Magnetism, Carnegie Institution
of Washington, 5241 Broad Branch Road NW, Washington, DC 20015-1305, USA}

\altaffiltext{6}{Department of Astronomy, University of California,
Berkeley, CA 94720, USA}

\altaffiltext{7}{UCO/Lick Observatory, University of California at Santa
Cruz, Santa Cruz, CA 95064, USA}

\begin{abstract} 
We obtained time-series spectroscopy of the G0 subgiant \eboo{} in an
attempt to confirm the solar-like oscillations reported by Kjeldsen et al.\
(1995).  We recorded 1843 spectra over six consecutive nights with the
Nordic Optical Telescope, which we used to measure equivalent widths of
strong temperature-sensitive lines.  We also measured velocities from 1989
spectra obtained through an iodine reference cell at Lick Observatory over
56 nights that were badly affected by weather.  Our analysis also included
velocity measurements published by Brown et al.\ (1997) and the original
equivalent-width measurements by Kjeldsen et al.\ (1995).  All four data
sets show power excesses consistent with oscillations, although with a
range of amplitudes that presumably reflects the stochastic nature of the
excitation.  The highest peaks show regularity with a large separation of
$\Dnu{}=40.4$\,\muHz{} and we identify 21 oscillation frequencies from the
combined data.  
\end{abstract}


\section{Introduction}

The search for solar-like oscillations is finally yielding success.
Observations of the sub-giants Procyon \citep{MSL99,BMM99} and
$\beta$~Hydri \citep{BBK2001,CBK2001} have shown very good evidence for
oscillations.  More recently, there has been an unambiguous detection of
p-mode oscillations in the main sequence star $\alpha$~Cen~A by
\citet{B+C2001,B+C2002}.  All these results were based on velocity
measurements obtained using high-dispersion spectrographs with stable
reference sources.

Another method for detecting oscillations was suggested by \citet[hereafter
Paper~I]{KBV95I}.  This involved monitoring changes in the equivalent
widths (EWs) of temperature-sensitive spectral lines.  In Paper~I we
reported evidence for oscillations in the G0 sub-giant $\eta$~Boo, based on
measurements of Balmer-line EWs.  We presented this as the first clear
evidence of solar-like oscillations in a star other than the Sun.  The
observations were obtained over six nights with the 2.5\,m Nordic Optical
Telescope on La Palma, and consisted of 12684 low-dispersion spectra.  In
the power spectrum of the equivalent-width measurements, we found an excess
of power at frequencies around 850\,\muHz.  The average amplitude inferred
for the oscillations was about 7 times greater than solar, in rough
agreement with the empirical scaling relation suggested by \citet{K+B95}.
Comb analysis of the power spectrum, described in Paper~I, suggested a
regular spacing of $\Dnu{}=40.3$\,\muHz.  Based on this, we identified
thirteen oscillation modes.  Similar observations of the daytime sky showed
the five-minute solar oscillations at the expected frequencies.

The frequencies for $\eta$~Boo reported in Paper~I, taken with available
estimates of the stellar parameters, were in good agreement with
theoretical models \citep{ChDBK95,G+D96}.  Particularly exciting was the
occurrence in theoretical models -- and apparently in the observations --
of `avoided crossings,' in which mode frequencies are shifted from their
usual regular spacing by effects of gravity modes in the stellar core.
Since then, the improved luminosity estimate for \eboo{} from Hipparcos
measurements has given even better agreement with the measured value of
$\Dnu{}$ \citep{BKChD98}.

Meanwhile, a search for velocity oscillations in \eboo{} by \citet{BKK97}
has failed to detect a signal, setting limits at a level below the value
expected on the basis of the EW results.  Although the data were sparse (22
hours spread over 7 successive nights) and the precision was degraded by
the relatively fast rotation of the star ($v\sin i = 13$\,km\,s$^{-1}$),
the analysis by \citet{BKK97} was careful and thorough, and the results
seem to be inconsistent with those of Paper~I.  More recently, \citet[][and
paper in preparation]{CBE2003} reported velocity measurements using the
CORALIE and ELODIE spectrographs that showed a clear excess of power and a
frequency spacing of $39.6$\,\muHz{}.

In this paper we present additional observations of \eboo, obtained in 1998
in both EW and velocity.  We also re-analyse our 1994 EW measurements and
the velocity measurements of \citet{BKK97}.  We confirm our earlier claim
for oscillations in \eboo{} and identify more than twenty oscillation
frequencies from the combined data.

\section{Data}

\subsection{Equivalent-width observations (NOT98)}

We observed \eboo{} over six nights during May 1998 using ALFOSC (Andalucia
Faint Object Spectrograph and Camera) on the 2.5\,m Nordic Optical
Telescope on La Palma.  This is the same telescope used in Paper~I but with
a different spectrograph.  We obtained seven echelle orders covering the
range 370--700\,nm at a dispersion of 0.04\,nm/pixel.
The CCD was a Loral 2k by 2k device, of which we used 700 by 1200 pixels.

Spectra were taken with a typical exposure time of 11\,s and a dead time
between exposures of 17\,s.  They were averaged in groups of three before
writing to disk, resulting in a total of 1843 spectra (sampling rate
1/84\,s) in 44.2 hours over six consecutive nights (1998 May 1--6).  The
distribution of spectra over the six nights was: 88, 222, 377, 391, 358 and
407.

\begin{figure*}
\epsscale{0.9} \plotone{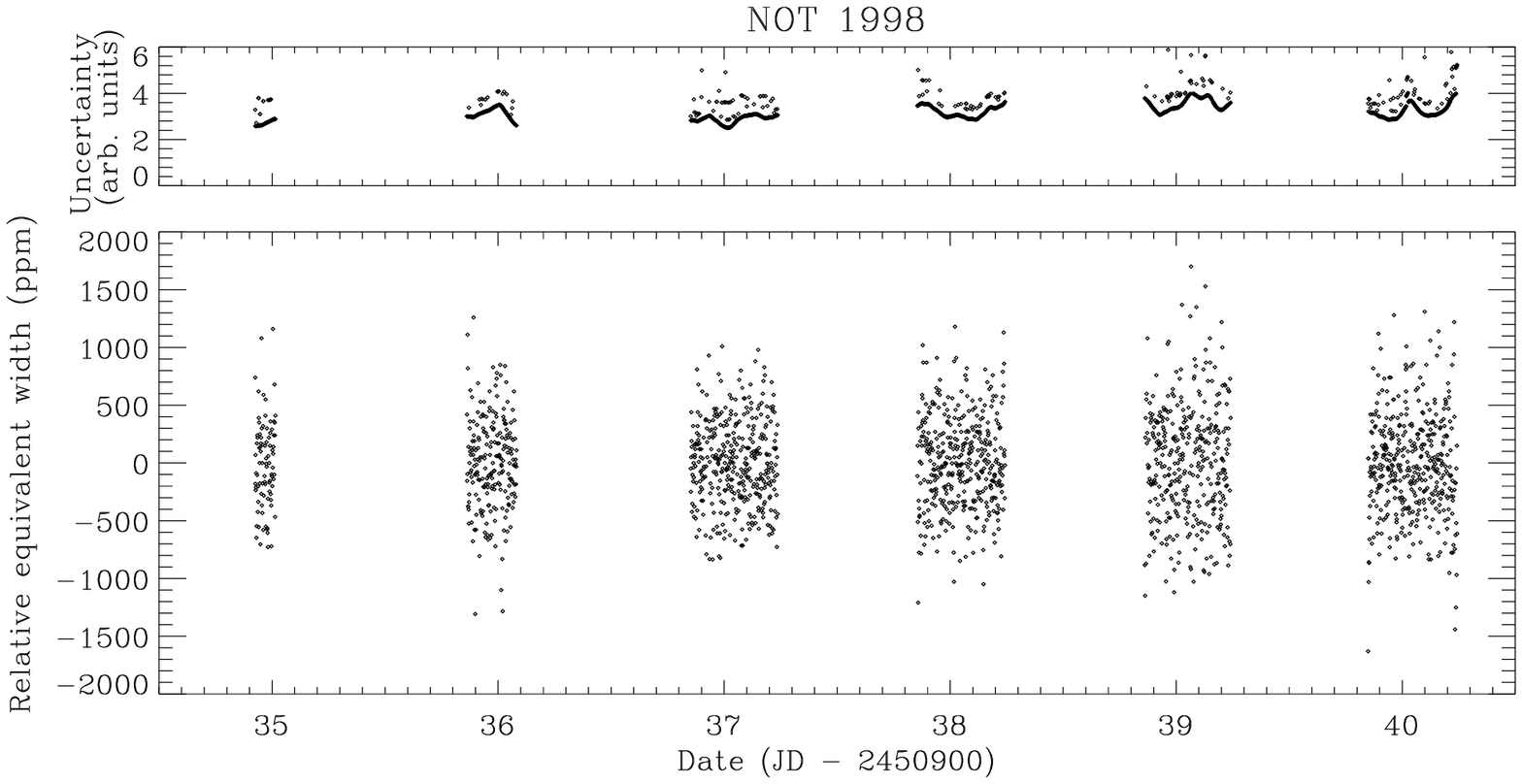}
\caption[]{\label{fig.not98-time}EW measurements of \eboo{}
obtained at the NOT 98 (lower panel) and the corresponding uncertainties
(upper panel).  }
\end{figure*}

We used the method described by \citet{KBF99} to measure equivalent widths
of six strong temperature-sensitive lines: H$\alpha$, H$\beta$, H$\gamma$,
Mg{\sc i}, Na{\sc i} and Fe{\sc i}.  A weighted mean of these six values
was calculated, taking into account the differing temperature sensitivities
of the lines.  The resulting time series is shown in
Fig.~\ref{fig.not98-time}.

\subsection{Velocity observations (Lick98)}

We also observed \eboo\ in 1998 with the Hamilton Echelle Spectrometer and
the 0.6-m Coud{\'e} Auxiliary Telescope (CAT) at Lick Observatory
\citep{Vog87}.  To produce high-precision velocity measurements, the star
was observed through an iodine absorption cell mounted directly in the
telescope beam.

We were allocated 56 of the 59 nights from 1998 April~6 to June~3, but the
weather was unseasonably poor, permitting observations on only 26 nights
(and many of these were partly lost).  The exposure time was 120\,s, with a
dead time between exposures of the same amount.  On the 11 best nights we
obtained 95--120 spectra per night (sampling rate: 1/245\,s), and the total
number of spectra obtained was 1989 (about one third of that possible with
no weather losses).

\begin{figure*}
\epsscale{0.9} \plotone{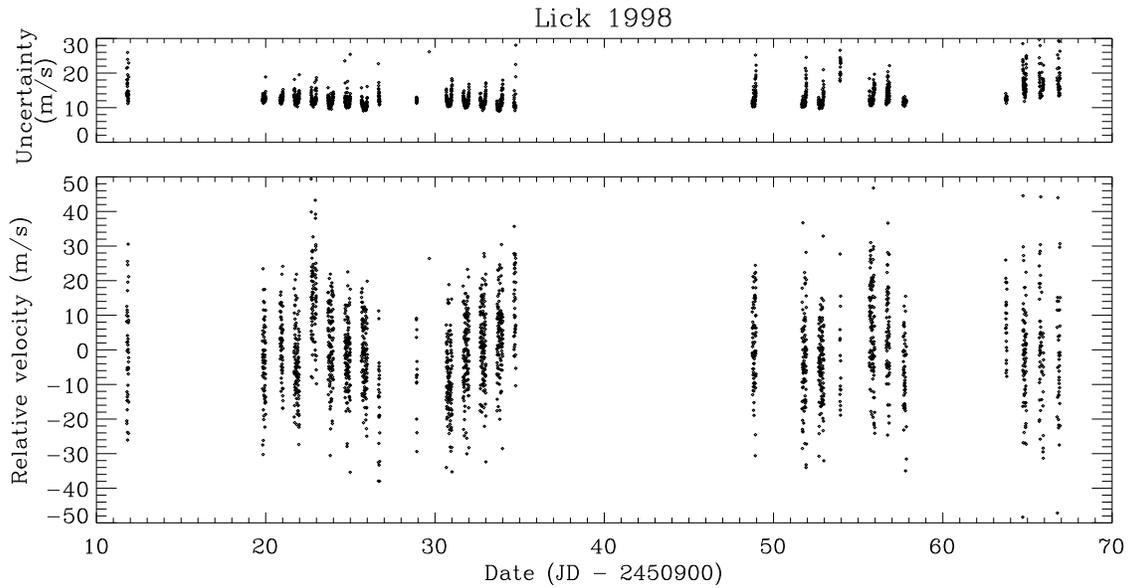}
\caption[]{\label{fig.lick-time}Velocity measurements of \eboo{}
obtained at Lick (lower panel) and the corresponding uncertainties
(upper panel).  }
\end{figure*}

Extraction of radial velocities from the echelle spectra followed the
method described by \citet{BMW96}.  As mentioned in the Introduction, the
precision is degraded by the relatively fast stellar rotation.  The star is
a spectroscopic binary with a period of 494\,d \citep{Ber57}, and the
orbital motion was removed from the velocity time series by fitting and
subtracting a fifth-order polynomial.  The resulting velocity measurements
are shown in Fig.~\ref{fig.lick-time}.  We are confident that the long-term
velocity variations are not instrumental, given that velocities for
$\tau$~Ceti, which we observed on most of the nights, were stable at the
5\,\ms{} level.  These night-to-night variations in \eboo{} are presumably
due to stellar activity, which is commonly observed in rotating G-type
stars \citep[see, e.g.,][]{SBM98,SMN2000}.

\subsection{Published equivalent-width observations (NOT94)}

We have included in this analysis the time series of 12684 EW measurements
obtained with the Nordic Optical Telescope in 1994.  These are identical to
the data presented in Paper~I, with the exception that a high-pass filter
was not applied.  The result is an increase in noise at low frequencies, as
expected from a $1/f$ noise source, which more accurately reflects the
actual stellar and instrumental noise \citep[for more details,
see][]{B+K95}.

\subsection{Published velocity observations (AFOE95)}

We have also analysed 555 velocity measurements of \eboo{} obtained with
the AFOE spectrograph during 22 hours spread over seven successive nights
in 1995 March.  These measurements were described by \citet{BKK97}
and were kindly provided to us in electronic form by Tim Brown.

\section{Analysis and discussion}  \label{sec.time-series}

\begin{figure*}
\epsscale{0.8} \plotone{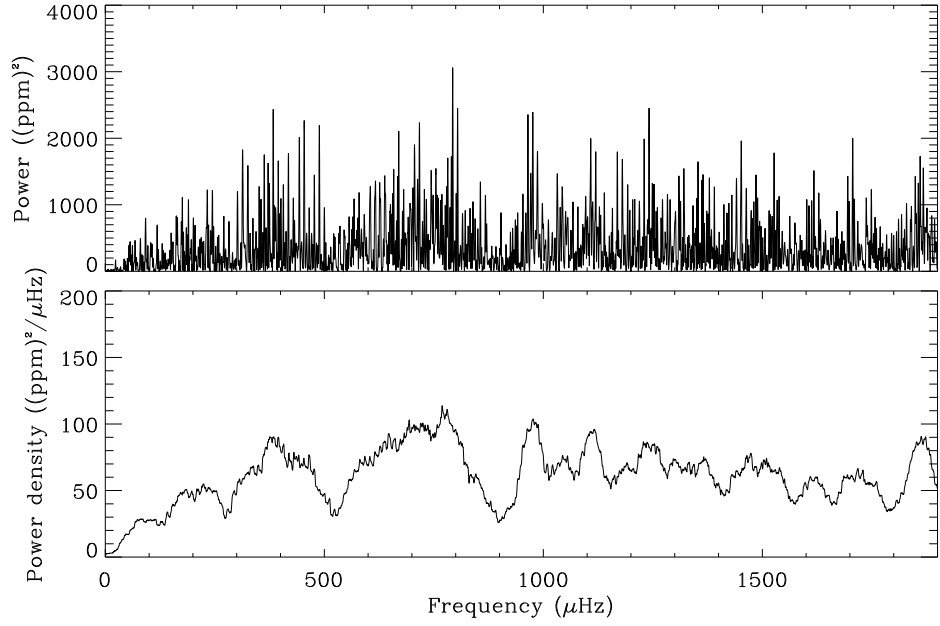}
\caption[]{\label{fig.not98-power} \label{fig.first-power} Power spectrum
of the NOT98 EW measurements of \eboo.  }
\end{figure*}

\begin{figure*}
\epsscale{0.8} \plotone{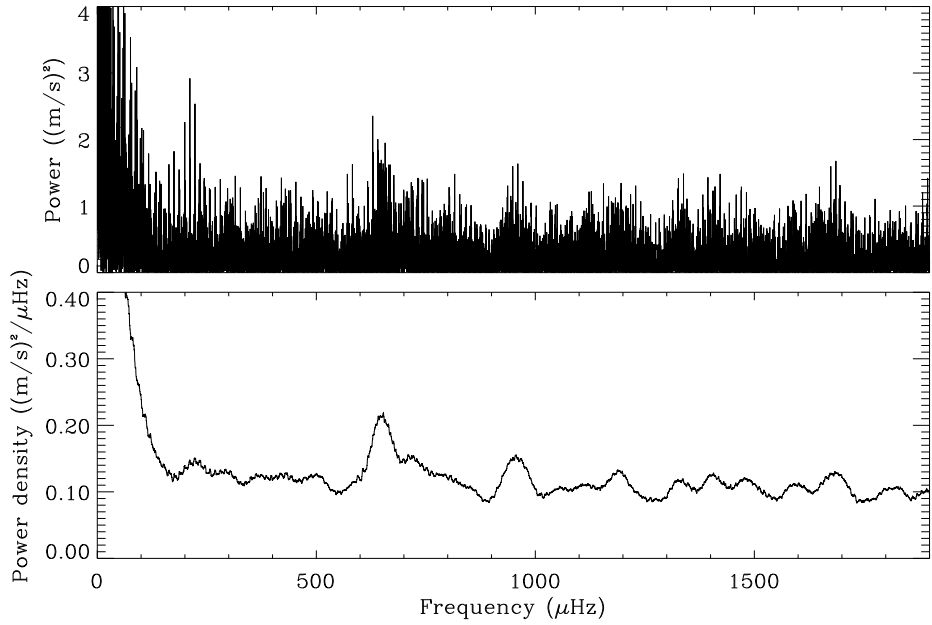}
\caption[]{\label{fig.lick-power} Power spectrum of the Lick98 velocity
measurements of \eboo{}.  }
\end{figure*}

\begin{figure*}
\epsscale{0.8} \plotone{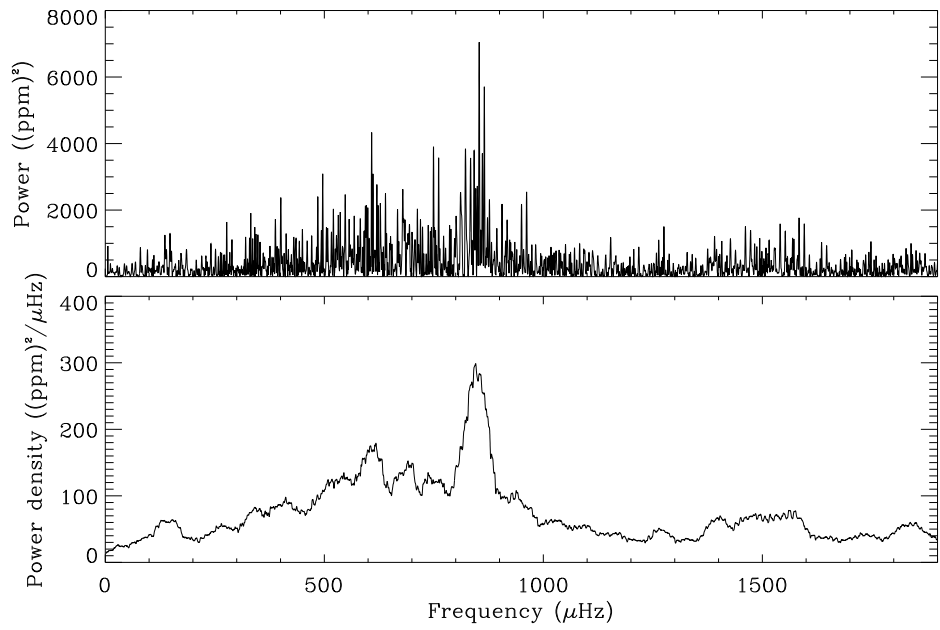}
\caption[]{\label{fig.not94-power} Power spectrum of the NOT94 EW
measurements of \eboo{} that were published in Paper~I.  }
\end{figure*}

\begin{figure*}
\epsscale{0.8} \plotone{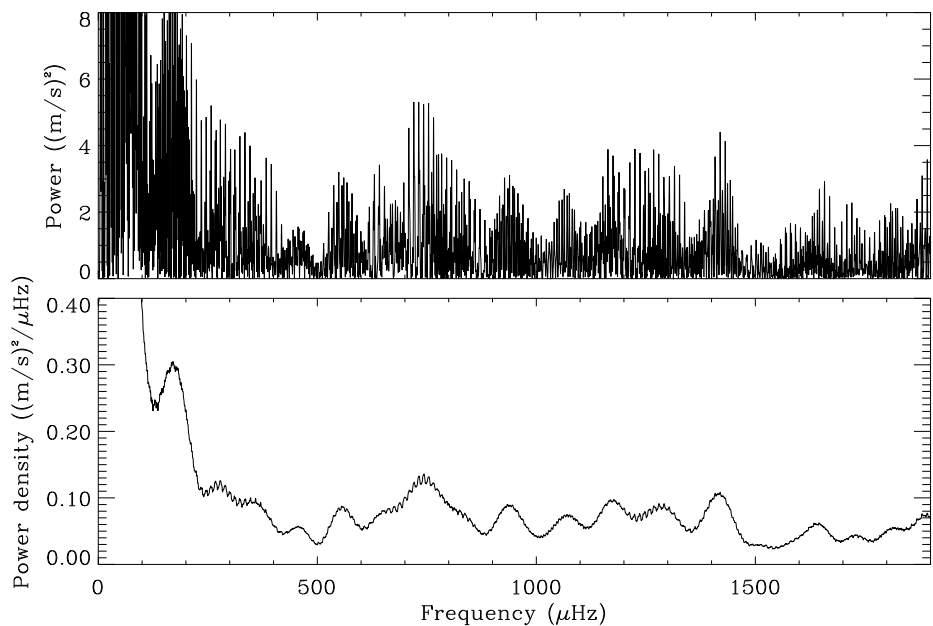}
\caption[]{\label{fig.afoe-power} \label{fig.last-power} Power spectrum
of the AFOE95 measurements of \eboo{} that were published by \citet{BKK97}.
}
\end{figure*}

The power spectrum of each time series was calculated as a weighted
least-squares fit of sinusoids \citep{FJK95,AKN98}, with a weight being
assigned to each point according its uncertainty estimate.  The results for
the four data sets are shown in
Figs.~\ref{fig.first-power}--\ref{fig.last-power}.  In each case, we show
both the conventional power spectrum (upper panels) and a smoothed version
in which the vertical scale has been converted to power density (lower
panels).  As discussed by \citet[Appendix A.1]{K+B95} and
\citet[Sec.~5]{KBF99}, the conversion to power density is achieved by
multiplying by the effective observing time, which we calculated in each
case by integrating under the spectral window.

Some level of excess power in the range 600--1100\,\muHz{} is apparent in
all four data sets, although it is strongest in the NOT94 data.  Further
discussion of the measured power levels is given below in
Sec.~\ref{sec.amplitudes}.  First, however, we will discuss the
oscillation frequencies.

Mode frequencies for low-degree solar-like oscillations are approximated
reasonably well by the asymptotic relation:
\begin{equation}
  \nu_{n,l} = \Dnu{} (n + \half l + \epsilon) - l(l+1) D_0.
        \label{eq.asymptotic}
\end{equation}
Here, $n$ and $l$ are integers that define the radial order and angular
degree of the mode, respectively; $\Dnu{}$ (the so-called large separation)
reflects the average stellar density, $D_0$ is sensitive to the sound speed
near the core and $\epsilon$ is sensitive to the surface layers.  It is
conventional to define $\dnu{02}$, the so-called small separation, as the
frequency spacing between adjacent modes with $l=0$ and $l=2$.  We can
further define $\dnu{01}$ to be the amount by which $l=1$ modes are offset
from the midpoint between the $l=0$ modes on either side.  If the
asymptotic relation holds exactly, then it is straightforward to show that
$D_0 = \sixth\dnu{02} = \half\dnu{01}$.

\subsection{Extraction of frequencies}

We extracted the frequencies of the strongest peaks in each power spectrum
in the range 600--1100\,\muHz.  We used a simple iterative algorithm, in
which the highest peak was identified and the corresponding sinusoidal
variation was subtracted from the time series.  The power spectrum of the
residuals was then calculated and the process was iterated until all peaks
with amplitudes more than 2.5 times the noise floor had been extracted.
The numbers of peaks extracted from the four data sets are summarized in
Table~\ref{tab.peaks}.

Setting the threshold at 2.5$\sigma$ gives us a chance to detect the weaker
oscillation modes, but it also means we will select some noise peaks.  To
investigate this, we have conducted simulations in which we analysed noise
spectra that contained no signal.  The last column in Table~\ref{tab.peaks}
shows the number of peaks found above the 2.5$\sigma$ threshold.  This
indicates that about 6 of the 35 detected peaks are expected to be due to
noise.

\subsection{Large frequency separation}

\begin{figure*}
\epsscale{0.5} \plotone{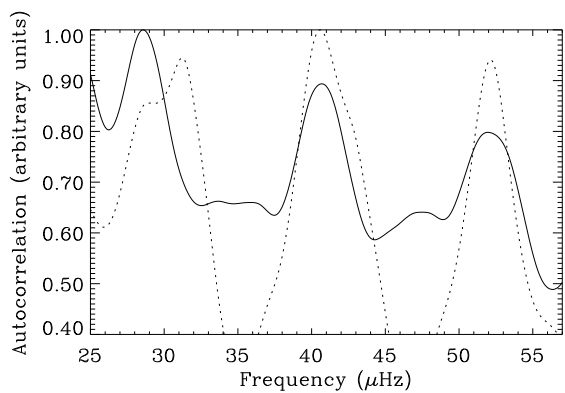}
\caption[]{\label{fig.auto} Solid line: the autocorrelation of the 22
frequencies, together with their daily aliases, that were extracted from
AFOE95, NOT98 and Lick98 data sets.  Dotted line: the same, but for the 13
frequencies from the NOT94 data.  The peaks at 40\,\muHz{} represent the
large frequency separation of \eboo.  }
\end{figure*}

We next investigated whether the extracted frequencies have a regular
spacing, as is expected for p-mode oscillations.  In Paper~I, we described
a comb analysis of the NOT94 power spectrum that revealed a regular spacing
of $\Dnu{}=40.3$\,\muHz.  Here, we analyse the 22 frequencies from the
other three data sets (AFOE95, NOT98, Lick98) using autocorrelation, as
follows.  Firstly, each extracted frequency was allocated a power
corresponding to the square of its S/N, in order to give higher weight to
the more reliable peaks.  Secondly, to allow for the likelihood that some
of the extracted frequencies should be shifted by $\pm$1\,d$^{-1}$
($\pm$11.57\,\muHz), we extended the table of frequencies by a factor of
three by including these side-lobes (but with half the power of the central
peaks).  We then calculated the autocorrelation of these 66 frequencies.
This is shown, smoothed to a resolution of 3.5\,\muHz, as the solid line in
Fig.~\ref{fig.auto}.

The three peaks in the autocorrelation
correspond to the large separation and its daily aliases, leading us to
estimate a value of $\Dnu{}=40.4$\,\muHz.  For comparison, the dotted line
in Fig.~\ref{fig.auto} shows the autocorrelation for the 13 frequencies
extracted from the NOT94 data, which yields a large separation of
$\Dnu{}=40.5$\,\muHz.  The excellent agreement between these independent
data sets confirms the result of Paper~I and also agrees well with the
value of $39.6$\,\muHz{} reported by \citet{CBE2003}.  In summary, there
can be no doubt that the large frequency separation of \eboo{} is about
40\,\muHz, which is in excellent agreement with theoretical models
\citep{ChDBK95,G+D96}.

\subsection{Identification of frequencies}

\begin{figure*}
\epsscale{0.8} \plotone{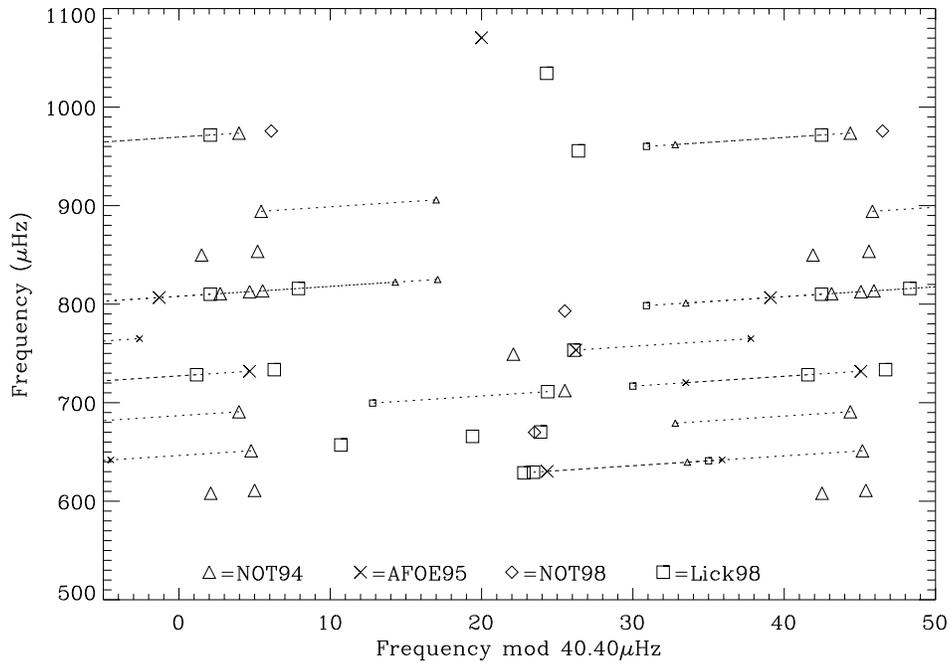}
\caption[]{\label{fig.echelle-raw} Echelle diagram showing the 35
frequencies from the four data sets.  The 15 cases for which we believe the
frequencies should be shifted by 1\,d$^{-1}$ (11.57\,\muHz) are represented
by two symbols connected by a dotted line, with the smaller symbol showing
the frequency without any shift.}
\end{figure*}

Given the large separation, we next attempted to identify the individual
modes, remembering that some of them will need to be shifted by
$\pm$11.57\,\muHz.  Figure~\ref{fig.echelle-raw} shows the 35 frequencies
from all four data sets, displayed in an echelle diagram.  We expect modes
with a given $l$~value to form vertical ridges, for $l=0$, $1$ and~$2$.  We
were able to achieve this by shifting 15 frequencies by $\pm$11.57\,\muHz,
as shown in the figure.  The final frequencies are given in
Table~\ref{tab.peaksid}.  Note that the number of peaks that are classified
as noise is consistent with the estimates made above (see
Table~\ref{tab.peaks}).

A possible problem is that, modulo the large separation, peaks
corresponding to ($\nu_{n,l} - 11.57$\,\muHz) for $l=1$ are only separated
by 1.1\,\muHz{} from peaks corresponding to ($\nu_{n,l} + 11.57$\,\muHz)
for $l=2$.  It is possible that some of those peaks have been shifted the
wrong way and are therefore wrongly identified.  The relevant peaks are
$699.6\,\muHz$ (from Lick98), which we identified as an alias of an $l=1$
mode, and $822.3\,\muHz$ (from NOT94), which we identified as an alias of
an $l=2$ mode.  A similar comment, although to a lesser extent, applies to
peaks corresponding to ($\nu_{n,l} - 11.57$\,\muHz) for $l=0$ and to
($\nu_{n,l} + 11.57$\,\muHz) for $l=1$, which are only separated by
1.9\,\muHz.

Note that we have identified the peak at 749.3\,\muHz{} as possibly being
an $l=1$ mode that is displaced by an avoided crossing.  This peak has the
second-highest S/N of all the peaks and is therefore not likely to be due
to noise or to be an alias.  In any case, shifting this peak by
$\pm$11.57\muHz{} would not bring it into agreement with $l=0$ or~2.

\begin{figure*}
\epsscale{0.8} \plotone{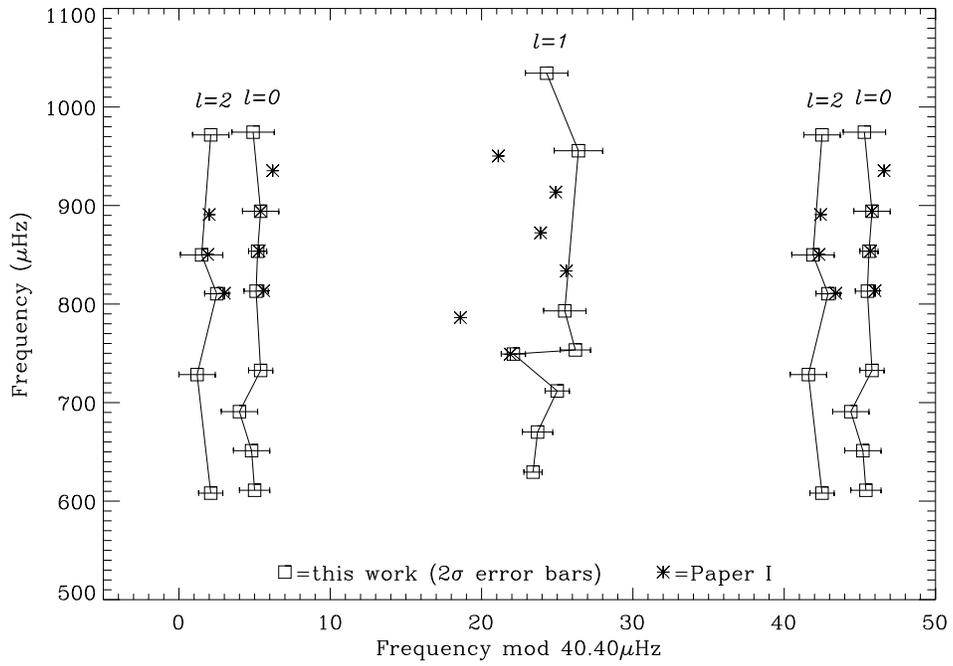}
\caption[]{\label{fig.echelle-final} Echelle diagram showing the 21
frequencies from Table~\ref{tab.final}, with $2\sigma$ error bars.  For
comparison, we also show the 13 frequencies reported in Paper~I.}
\end{figure*}

Some of the frequencies are detected more than once, as indicated by the
ditto marks~('') in Table~\ref{tab.peaksid}.  In those cases, we have
combined the measurements into a weighted average.  The final list of 21
frequencies is given in Table~\ref{tab.final}, and these are shown as an
echelle diagram in Fig.~\ref{fig.echelle-final}.  The uncertainties in the
frequencies reflect the S/N of the relevant peak (or peaks).  

We next estimated the large separation separately for each value of $l$,
and the results are given in the last line of Table~\ref{tab.final} (the
two modes marked with an asterisk, which may be affected by avoided
crossings, were excluded from the fit).  The weighted average of these
three $\Dnu{l}$ yields the value of $\Dnu{}$ shown in
Table~\ref{tab.splittings}.  The other parameters in that table were
calculated by fitting the same 19 frequencies to the asymptotic relation.
All are consistent with the values given in Paper~I but have smaller
uncertainties, thanks to the larger number of detected frequencies.

In Fig.~\ref{fig.echelle-final} we also show the 13 frequencies reported in
Paper~I.  We can see that six were recovered in the new analysis, while
five were not recovered but lie close to one of the ridges (within typical
uncertainties), so perhaps can still be taken as reliable detections.
Finally, two frequencies (786.2 and 950.3\,\muHz) fall well away from the
$l=1$ ridge and were not confirmed by the revised analysis.  It is possible
that they represent mixed modes that are shifted by avoided crossings, but
further observations are needed to confirm this.

\subsection{Oscillation amplitudes} \label{sec.amplitudes}

The above analysis gives strong evidence that the highest peaks in the four
power spectra are due to solar-like oscillations.  However, we have also
noted that the amplitude of the power excess is somewhat stronger in the
NOT94 observations than in the other three data sets.  It seems plausible
that these differences reflect the stochastic nature of the excitation
mechanism, but a more definite statement is hampered by our limited
knowledge of the amplitudes of oscillations in subgiant stars.  It is also
possible that some contribution to the variations is due to an unidentified
systematic error in the calibration that was used to convert
equivalent-width amplitudes to velocities (see Paper~I).

Taken at face value, the observations indicate that peak oscillation
amplitudes in \eboo{} are typically 3--5 times solar.  This conclusion is
consistent with the upper limits reported by \citet{BKK97} from the AFOE95
data.

\section{Conclusions}

We have presented new observations of \eboo{} in velocity (Lick98) and
equivalent width (NOT98) that show some evidence for excess power in the
range 600--1100\,\muHz.  The velocity measurements published by
\citet[][AFOE95]{BKK97} also show a slight power excess when smoothed.
None of these signals is as strong as the original equivalent-width
observations (Paper~I, NOT94), which may reflect the stochastic nature of
the excitation or may indicate a problem with the calibration of the NOT94
equivalent-width estimates.

We extracted the highest peaks in each data set and used the
autocorrelation to search for regularity.  The three newer data sets
(AFOE95, NOT98 and Lick98) combined to give a clear autocorrelation signal
at a spacing of $\Dnu{}=40.4$\,\muHz, giving independent confirmation of
the result of Paper~I.  This, combined with the recent work by
\citet{CBE2003}, leaves little doubt that the large frequency separation of
\eboo{} is about 40\,\muHz.  This allowed us to identify 21 frequencies in
\eboo{} which have been compared with theoretical models by
\citet{DiMChDK2003}.  The results confirm the claim made in Paper~I for the
first clear evidence of solar-like oscillations in a star other than the
Sun.  Future observations of \eboo, particularly with the {\em MOST\/}
spacecraft \citep{MKW2000}, should measure more oscillation modes with
greater frequency precision.

We would be pleased to make the data available on request.  Please contact
Tim Bedding ({\small\verb"bedding@physics.usyd.edu.au"}).

\acknowledgments

We thank Tim Brown for providing the AFOE velocity measurements of \eboo.
This work was supported financially by the Australian Research Council
(TRB and IKB), National Science Foundation Grant AST-9988087 (RPB), SUN
Microsystems, and the Danish Natural Science Research Council and the
Danish National Research Foundation through its establishment of the
Theoretical Astrophysics Center (HK).

\clearpage

\clearpage

\clearpage

\begin{table*}
\small
\caption{\label{tab.peaks} Results of the frequency analysis}
\begin{center}
\begin{tabular}{cccc}
\tableline
\tableline
\noalign{\smallskip}
Dataset &  Noise level &  \multicolumn{1}{c}{Extracted peaks} & Noise peaks\\
\tableline
\noalign{\smallskip}
NOT94   & 15.1 ppm  &    13 & 1.3 $\pm$ 0.4 \\
AFOE95  & 0.67 \ms  &   ~~5 & 1.0 $\pm$ 0.3 \\
NOT98   & 16.3 ppm  &   ~~3 & 0.5 $\pm$ 0.2 \\
Lick98  & 0.41 \ms  &    14 & 3.3 $\pm$ 1.0 \\
\tableline
\end{tabular}
\end{center}
\end{table*}

\begin{table*}
\small
\caption{\label{tab.peaksid} Identification of extracted frequencies}
\begin{center}
\begin{tabular}{rccc}
\tableline
\tableline
\noalign{\smallskip}
\multicolumn{1}{c}{Frequency} & Mode ID & Dataset &      S/N  \\
\multicolumn{1}{c}{(\muHz)}   &         &         &           \\
\tableline              
\noalign{\smallskip}
    $ 608.1                 $ &   $l=2$ & NOT94   &    $4.4$  \\
    $ 611.0                 $ &   $l=0$ & NOT94   &    $3.6$  \\
    $ 628.8                 $ &   $l=1$ & Lick98  &    $3.7$  \\
    $ 641.0-11.57 = 629.4   $ &    ''   & Lick98  &    $3.0$  \\
    $ 641.9-11.57 = 630.3   $ &    ''   & AFOE95  &    $3.0$  \\
    $ 639.6+11.57 = 651.2   $ &   $l=0$ & NOT94   &    $3.1$  \\
    $ 657.1                 $ &   noise & Lick98  &    $3.2$  \\
    $ 665.8                 $ &   noise & Lick98  &    $2.9$  \\
    $ 669.9                 $ &   $l=1$ & NOT98   &    $2.7$  \\
    $ 670.3                 $ &    ''   & Lick98  &    $2.7$  \\
    $ 679.2+11.57 = 690.8   $ &   $l=0$ & NOT94   &    $3.1$  \\
    $ 699.6+11.57 = 711.2   $ &   $l=1$ & Lick98  &    $2.9$  \\
    $ 712.3                 $ &    ''   & NOT94   &    $3.3$  \\
    $ 716.8+11.57 = 728.4   $ &   $l=2$ & Lick98  &    $3.0$  \\
    $ 720.3+11.57 = 731.9   $ &   $l=0$ & AFOE95  &    $3.4$  \\
    $ 733.5                 $ &    ''   & Lick98  &    $2.9$  \\
    $ 749.3                 $ &   $l=1$?& NOT94   &    $4.4$  \\
    $ 753.3                 $ &   $l=1$ & Lick98  &    $2.6$  \\
    $ 765.0-11.57 = 753.4   $ &    ''   & AFOE95  &    $2.9$  \\
    $ 793.1                 $ &   $l=1$ & NOT98   &    $2.8$  \\
    $ 806.7                 $ &   noise & AFOE95  &    $3.0$  \\
    $ 798.5+11.57 = 810.1   $ &   $l=2$ & Lick98  &    $2.6$  \\
    $ 822.3-11.57 = 810.7   $ &    ''   & NOT94   &    $4.3$  \\
    $ 801.1+11.57 = 812.7   $ &   $l=0$ & NOT94   &    $3.1$  \\
    $ 825.1-11.57 = 813.5   $ &    ''   & NOT94   &    $3.1$  \\
    $ 815.9                 $ &   noise & Lick98  &    $3.0$  \\
    $ 849.9                 $ &   $l=2$ & NOT94   &    $2.8$  \\
    $ 853.6                 $ &   $l=0$ & NOT94   &    $5.6$  \\
    $ 905.8-11.57 = 894.2   $ &   $l=0$ & NOT94   &    $3.2$  \\
    $ 955.6                 $ &   $l=1$ & Lick98  &    $2.5$  \\
    $ 960.1+11.57 = 971.7   $ &   $l=2$ & Lick98  &    $3.0$  \\
    $ 962.0+11.57 = 973.6   $ &   $l=0$ & NOT94   &    $3.2$  \\
    $ 975.7                 $ &    ''   & NOT98   &    $2.9$  \\
    $1034.3                 $ &   $l=1$ & Lick98  &    $2.6$  \\
    $1070.4                 $ &   noise & AFOE95  &    $2.5$  \\
\tableline
\end{tabular}                                                 
\end{center}
\end{table*}

\begin{table*}
\small
\caption{\label{tab.final} Oscillation frequencies for \eboo{} (\muHz)}
\begin{center}
\begin{tabular}{crrr}
\tableline
\tableline
\noalign{\smallskip}
 & \multicolumn{1}{c}{$l=0$}  & \multicolumn{1}{c}{$l=1$} & \multicolumn{1}{c}{$l=2$}\\
\tableline
\noalign{\smallskip}
$n=13$ &             &              & 608.1 (0.4) \\
$n=14$ & 611.0 (0.5) &  629.4 (0.3) &             \\
$n=15$ & 651.2 (0.6) &  670.1 (0.5) &             \\
$n=16$ & 690.8 (0.6) &  711.8 (0.4) & 728.4 (0.6) \\
$n=17$ & 732.6 (0.4) &  749.3 (0.4)\rlap{$^*$} &             \\
$    $ &             &  753.4 (0.5)\rlap{$^*$} &             \\
$n=18$ &             &  793.1 (0.7) & 810.5 (0.4) \\
$n=19$ & 813.1 (0.4) &              & 849.9 (0.7) \\
$n=20$ & 853.6 (0.3) &              &             \\
$n=21$ & 894.2 (0.6) &              &             \\
$n=22$ &             &  955.6 (0.8) & 971.7 (0.6) \\
$n=23$ & 974.5 (0.7) &              &             \\
$n=24$ &             & 1034.3 (0.7) &             \\
\noalign{\bigskip}
$\Dnu{l}$&40.45 (0.07) & 40.89 (0.19) & 40.41 (0.10) \\
\tableline
\end{tabular}
\end{center}
\end{table*}

\begin{table*}
\small
\caption{\label{tab.splittings} Frequency separations for \eboo{}}
\begin{center}
\begin{tabular}{cr}
\tableline
\tableline
\noalign{\smallskip}
$\Dnu{}  $	& $40.47 \pm 0.05\,\muHz$\\
$\dnu{02}$ 	& $3.00 \pm 0.35\,\muHz$\\
$\dnu{01}$ 	& $0.78 \pm 0.45\,\muHz$\\
$D_0 	 $	& $0.49 \pm 0.06\,\muHz$\\
$\epsilon$ 	& $1.09 \pm 0.02\phantom{\,\muHz}$\\
\tableline
\end{tabular}
\end{center}
\end{table*}

\end{document}